# Defect trapping and phase separation in chemically doped bulk AgF$_2$

Adam Grzelak[1]*, Mariana Derzsi[1,2], Wojciech Grochala[1]*

[1]*Center of New Technologies, University of Warsaw, Banacha 2C, 02-097 Warsaw, Poland*
[2]*Advanced Technologies Research Institute, Faculty of Materials Science and Technology in Trnava, Slovak University of Technology in Bratislava, Jána Bottu 8857/25, 917 24 Trnava, Slovakia*

*This work is dedicated to prof. Boris Žemva at his 80$^{th}$ birthday*

ABSTRACT

We report a computational survey of chemical doping of silver(II) fluoride, an oxocuprate analog. We find that the ground-state solutions exhibit strong tendency for localization of defects and for phase separation. The additional electronic states are strongly localized and the resulting doped phases exhibit insulating properties. Our results, together with previous insight from experimental attempts, indicate that chemical doping may not be a feasible way towards high-temperature superconductivity in bulk silver(II) fluoride.

INTRODUCTION

Silver(II) fluoride has recently attracted attention of chemists and physicists alike because of its similarities to precursors of copper oxide-based superconductors. [1–3] Charge doping within their constituent Cu-O$_2$ layers is a key step towards achievement of superconductivity in cuprates. [4] Given the aforementioned similarities, it is justified that doping of AgF$_2$ should also be studied.

Charge doping of a compound can be achieved experimentally by several different means. For example, using field-effect transistor setup it is possible to alter the concentration of charge carriers to a desired degree and produce materials with modified electronic and magnetic properties. [5] At the computational level, this can be simulated e.g. by explicitly setting the number of electrons in the unit cell to a number that differs from the sum of all valence electrons of its constituent atoms. This has already been attempted by our group in bulk AgF$_2$ and in hypothetical monolayers. [6] The latter, although as yet unknown, could in principle be obtained by epitaxy. [7] Another approach is to synthesize a material with a modified (e.g. non-stoichiometric) composition, leading to mixed (or intermediate) valence of one of its constituent elements. That is the case with copper oxide-based superconductors: for example, in La$_2$CuO$_4$, the first known precursor to high-temperature superconductors, a substitution of a fraction of La(III) cations for e.g. Ba(II) [8] or an increase of nonmetal (content [9] leads to hole-doping of Cu-O layers. Computationally, this can be easily modelled by inserting, removing or substituting certain atoms in the studied unit cell. In this contribution, we utilize a similar approach. By changing the number of F atoms in the AgF$_2$ unit cell, we obtain different levels of charge doping and analyze crystal, electronic and magnetic structures of the resulting polymorphs. This in turn enables us to assess the prospect for cuprate-like properties in these materials.

COMPUTATIONAL METHODS

All calculations were performed with density functional theory (DFT) approach as implemented in VASP software [10–14], using GGA-type Perdew-Burke-Ernzerhof functional adapted for solids (PBEsol). [15]

Plane-wave cutoff energy of 520 eV was used. k-spacing of 0.04 Å$^{-1}$ was used for structural optimization and 0.02 Å$^{-1}$ for electronic density of states (eDOS) calculations. For spin-polarized calculations, on-site Coulombic interactions of Ag d electrons were accounted for through DFT+U correction as introduced by Liechtenstein *et al.*, [16] with the Hubbard *U* and Hund $J_H$ parameters for Ag set to 5 eV and 1 eV, respectively. [17] VESTA software was used for visualization of crystal structures. [18] Electronic density of states (eDOS) and local potential graphs were plotted with p4vasp software. [19]

RESULTS AND DISCUSSION

1. CRYSTAL AND MAGNETIC STRUCTURE

It should be noted that $AgF_2$, although similar in many ways to $La_2CuO_4$, [1] is a binary [001] compound, which greatly reduces the number of way its stoichiometry can be modified, as compared to ternary $La_2CuO_4$. Our starting point was a 2x2x2 supercell of silver(II) fluoride (which crystallizes in a *Pbca* orthorhombic structure [20]) containing 32 formula units (FU) of $AgF_2$, pre-optimized using GGA+U approach described in the previous section. We then modified the supercell by removing or adding fluorine atoms into the structure, which led to electron or hole doping, respectively. We investigated three doping levels on each side of the phase diagram: 1/32, 1/16 and 1/8, which correspond to removal/addition of 2, 4 and 8 F atoms, respectively, compared to the starting supercell. The optimization of modified supercells was carried out in two steps. In the first step, the structures were optimized at GGA level, without Hubbard correction and magnetic interactions. After that, magnetic models were constructed in pre-optimized structures, taking into account changes in local coordination of Ag sites and intuitions from Goodenough-Kanamori-Anderson rules. [21] These magnetic models were taken as starting point for optimization at GGA+U level, with spin polarization and Hubbard correction described in the previous section. Several possible patterns of removal/addition of F atoms were studied at each doping level; only the lowest-energy solutions are presented and discussed in this work. Their structures are shown in fig. 1.

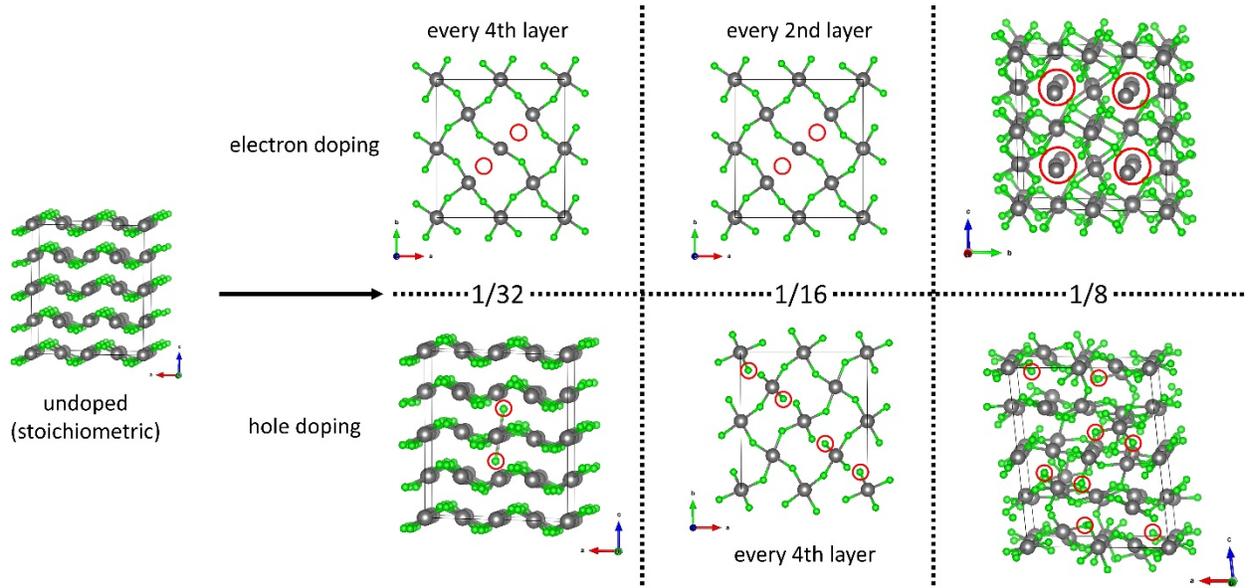

Fig. 1. Structures of modified $AgF_2$ supercells. Fractions represent doping level. Red circles indicate vacancies and interstitial for electron and hole doping, respectively. For e-doped 1/8 solution, emptied local environment of several Ag atoms is highlighted instead for clarity. "Every *n*-th layer" indicates that this pattern appears in 4/*n* out of 4 layers in the unit cell.

Ground-state structure of bulk stoichiometric AgF$_2$ consists of layers stacked along the *c* dimension of the unit cell. The most stable solutions of doped AgF$_2$ investigated in this work exhibit some common features, which point to overall tendencies associated with doping. The lowest-energy solutions of 1/32 and 1/16 e-doped models are both characteristic of the same pattern of F vacancies aligned along the *ab* diagonal of the unit cell, only in the 1/16 solution the pattern is repeated every two layers along the *c* direction and shifted by (0.5,0.5) in the *ab* plane. Also, in the hole-doped solutions at 1/32 and 1/16 level, the defects – in this case additional F atoms – are localized around a particular Ag site or contained within a single layer. In contrast, at the 1/8 level, a more complex, three-dimensional pattern emerges in both e-doped and h-doped solution.

In order to understand the influence of fluorine vacancies and interstitials, we have to take a closer look at their magnetic structure (fig. 2). In bulk AgF$_2$ (not shown), each layer within the *ab* plane features two-dimensional antiferromagnetic (AFM) ordering of spins on Ag(II) sites (with magnetic moment of ca. 0.57 µ$_B$). In the first approximation and assuming localization of additional charge, the number of new Ag(I) cations in e-doped solutions should be equal to the number of F vacancies (conversely, the same applies to number of Ag(III) cations and F interstitials in h-doped solutions). We find it to be true for solutions at 1/8 doping level. In e-doping, Ag(I) cations are located in channels along the *a* directions within a three-dimensional framework of AgF$_2$. There is also a lot of additional spin (ca. 0.10 µ$_B$) on dangling F atoms, i.e. those connected only to one Ag(II) cation; magnetic moments of this size are in fact typical for fluoroconnections of Ag(II). [22] Ag(I)-F contacts are not shown for clarity, but they are also substantially longer than Ag(II)-F: 2.3-2.6 Å compared to 2.0-2.1 Å (more detailed description of local coordination will be presented below). In h-doped 1/8 solution, there are eight clearly discernible Ag(III) sites: two of high-spin d$^8$ electronic configuration – with high magnetic moment (ca. 0.86 µ$_B$) and roughly octahedral coordination – and six of low-spin configuration, with null magnetic moment and roughly square-planar coordination.

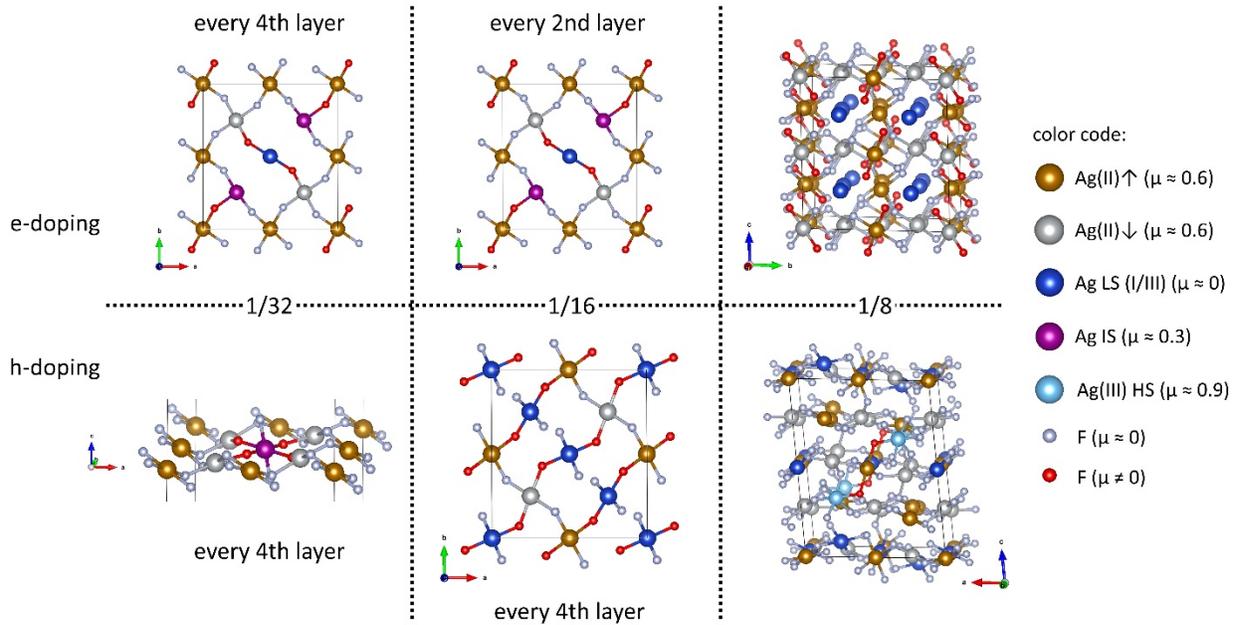

Fig. 2. Magnetic ordering in ground-state doped AgF$_2$ solutions. Fractions represent doping level. "µ" refers to magnetic moment in units of µ$_B$. LS – low-spin, HS – high-spin, IS – intermediate-spin. "Every *n*-th layer" indicates that this pattern appears in 4/*n* out of 4 layers in the unit cell.

In e-doped solutions at 1/32 and 1/16 levels, there is a clearly discernible Ag(I) site in the middle of the defect pattern, with two additional Ag sites of intermediate spin (0.30 $\mu_B$, i.e. between null and the value typical for regular Ag(II)), which we also designated as contributing to Ag(I) states in their electronic structure (discussed in the corresponding section later in the text). Most probably, such distribution of spin is a superposition of two solutions, in either of which a second Ag(I) cation occupies only one of those two positions. However, such arrangement would destroy the center of inversion at the central Ag(I) site and therefore does not arise. The inversion center at (0.5,0.5,0.5) is the only symmetry constraint in our models, *which we consider a reasonable compromise between flexibility of geometry optimization and use of computational resources*. Finally, h-doped solution at 1/32 level features the hole localized on a single central Ag site and its neighboring F atoms, while the 1/16 solution hosts four low-spin Ag(III) cations confined to a single $AgF_2$ layer.

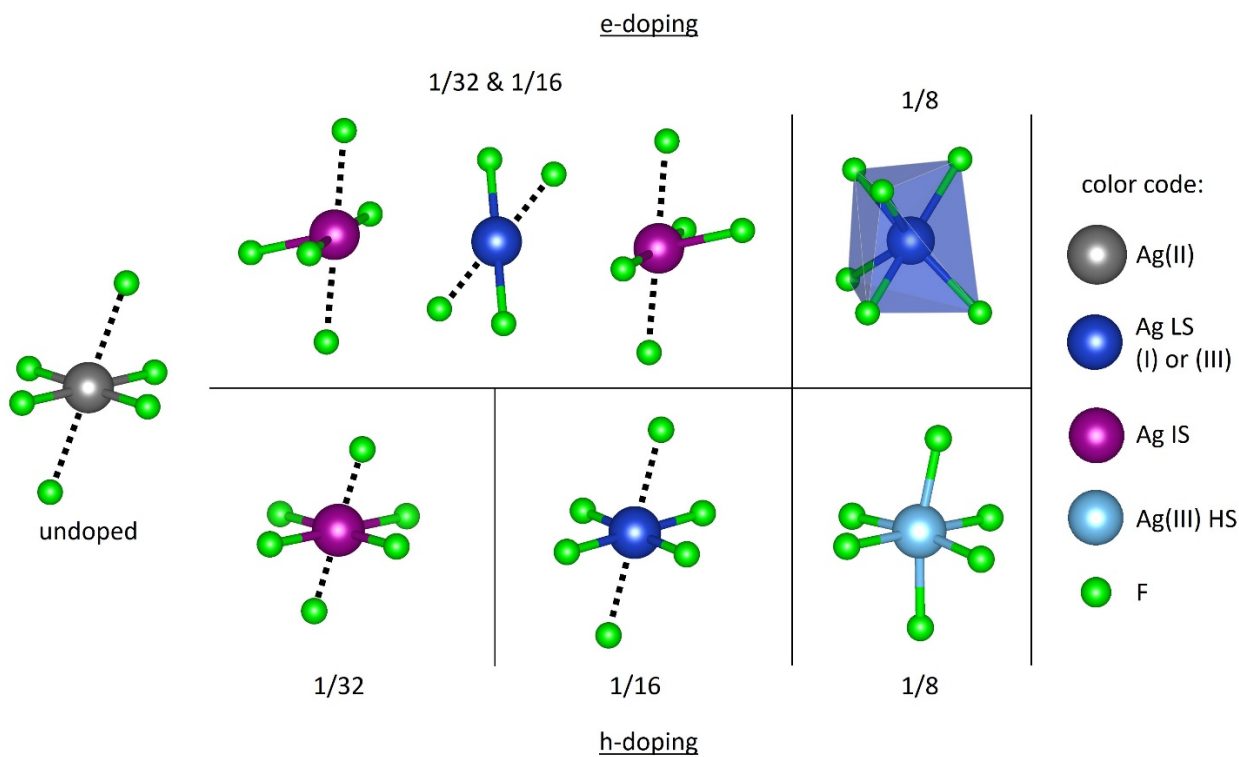

Fig. 3. Comparison of local coordination of an Ag(II) site in undoped $AgF_2$ and of Ag(I) and Ag(III) sites in doped solutions. LS – low-spin, HS – high-spin, IS – intermediate-spin.

Further insight into the influence of doping on crystal structure of $AgF_2$ is provided through examination of local coordination of Ag sites (fig. 3 and table 1). In stoichiometric (undoped) $AgF_2$, all Ag sites exhibit a coordination pattern of 4+2 F atoms in an elongated, tilted octahedron, characteristic of Jahn-Teller-active $d^9$ metal cations. In doped solutions, Ag centers with altered magnetic moment undergo a concomitant change in local coordination. The Ag(I) species in 1/32 and 1/16 e-doped solutions show an increased length of the two in-plane F contacts, with simultaneous shortening of the axial (inter-layer) F contacts. In the 1/8 solution, the Ag(I) cation is surrounded by 6 or 7 F atoms at a length between ca. 2.3-2.6 Å. (Note: only the six-fold coordination in trigonal prism pattern is shown in fig. 3; half of Ag(I) cations in this solution also picks up a seventh F atoms to form a trigonal prism monocapped on one of its rectangular faces.) Such changes point to a less directional and more spherical distribution of electrons, and disappearance of the first-order Jahn-Teller effect, which is of course expected for a closed-shell ($d^{10}$) Ag(I) cation. In addition, the average length of F contacts is slightly larger than for Ag(II) species – a further evidence of decreased oxidation state and stronger repulsion between atoms. These effects can, to some extent, be observed also for the intermediate-spin Ag sites in e-doped solutions at 1/32 and 1/16 level.

On the other hand, in h-doped solutions, an approximately octahedral coordination is retained also for Ag(III) and intermediate-spin Ag sites. The former – both the high-spin and low-spin species – show an overall decrease of average Ag-F bond length (compared to Ag(II)), which is expected at their higher oxidation state. The intermediate-spin Ag sites are coordinated in a pattern of a compressed tilted octahedron. The shortened Ag-F axial bonds can be explained by a) the fact that these F atoms are only bonded to one Ag atom and therefore their electron cloud is redistributed to minimize repulsion, b) their non-zero magnetic moment, indicating the partial presence of holes and thus decreased electron density.

Table 1. Comparison of local coordination of Ag sites – numerical data. Distances are given in angstroms. Numbers next to species name (in bold) designate coordination numbers. LS – low-spin, HS – high-spin, IS – intermediate-spin. Footnote: distances in the table are averaged over the instances of that type of Ag-F contact at that particular type of site and, in some cases (indicated with †), over more than one solution in which that type of site appears.

| undoped AgF$_2$ | | electron doping | | | | | |
|---|---|---|---|---|---|---|---|
| **Ag(II) – 4+2** | | **Ag(I) – 2+2†** | | **Ag IS – 3+2†** | | **Ag(I) – 6(+1)** | |
| square | axial | line | axial | in-plane | axial | trigonal prism (monocapped) | |
| 2.069 | 2.570 | 2.151 | 2.408 | 2.189 | 2.408 | 2.431 | |
| | | hole doping | | | | | |
| | | **Ag IS – 4+2** | | **Ag(III) LS – 4+2†** | | **Ag(III) HS – 6** | |
| | | square | axial | square | axial | octahedron | |
| | | 2.035 | 1.884 | 1.931 | 2.509 | 2.056 | |

## 2. DEFECT CLUSTERING AND PHASE SEPARATION

Joint analysis of structural and magnetic data points to a strong tendency of bulk AgF$_2$ for localization of defects resulting from doping (fig. 4). In most stable e-doped solutions, the additional charge tends to localize at particular atoms and is not distributed evenly across the unit cell. In addition, these defects can also be clustered close to each other within a layer of AgF$_2$ – for example, at 1/16 level of h-doping, all defects in the supercell are clustered within one out of four layers. This effectively leads to a structure in which 3 layers are undoped and one of them is 1/4-h-doped. This may indicate that such system would in fact be unstable towards phase separation, according to the equation:

Eq. (1) 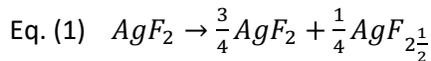

$$AgF_2 \rightarrow \frac{3}{4} AgF_2 + \frac{1}{4} AgF_{2\frac{1}{2}}$$

where AgF$_{2½}$ corresponds to the known fluoride of [Ag(II)F][Ag(III)F$_4$] stoichiometry. [23]

As can be seen in fig. 5, the free energy of formation of doped polymorphs is positive in almost all cases, with the exception of h-doping at 1/8 level, where it is only slightly negative – equal to –0.02 eV. (Note: free energy is in this instance equivalent to enthalpy, as the calculations are performed at T → 0 K, so the entropy term is null, and at constant pressure, formally p → 0 GPa.)

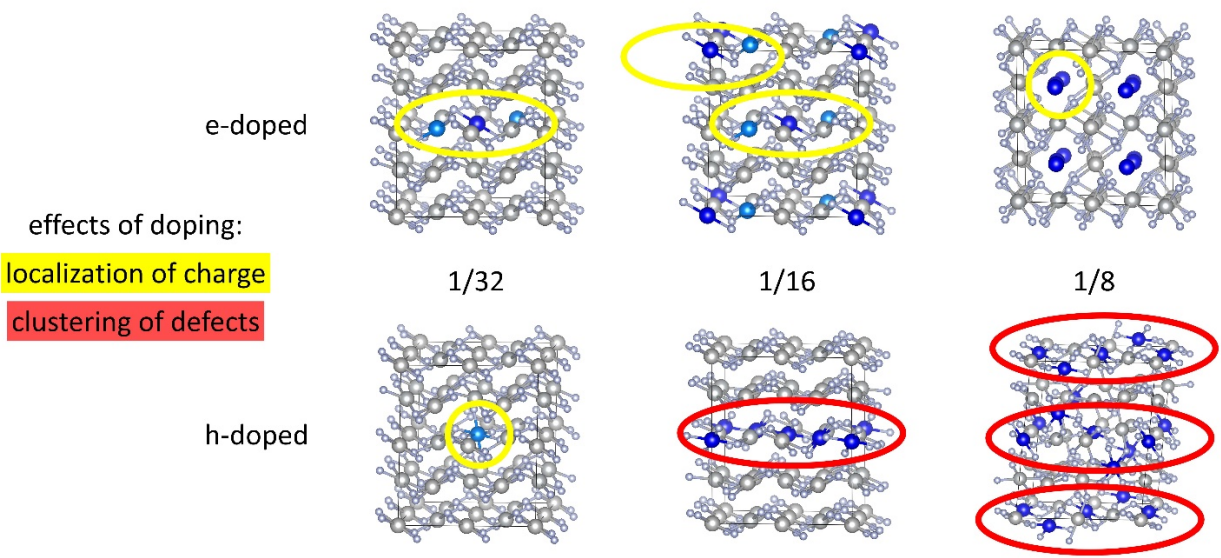

Fig. 4. Illustration of charge localization and clustering of defects in doped AgF$_2$ polymorphs. Large grey spheres – Ag(II), blue spheres – Ag(I)/Ag(III), small grey spheres – F.

The generalized reaction equations, which reflect doping, are:

Eq. (2)  $AgF_2 \rightarrow AgF_{2(1-x)} + xF_2$     (electron doping)

and

Eq. (3)  $AgF_2 + xF_2 \rightarrow AgF_{2(1+x)}$     (hole doping)

Therefore, for electron doping, the equilibrium would be shifted towards products at higher temperatures, as it would increase the entropy of gaseous fluorine. According to our estimates, which are based on our results and thermochemical data for fluorine [24], temperatures required for synthesis of e-doped polymorphs would amount to *ca.* 1300 K for 1/8 doping level and ca. 1500 K for 1/16 and 1/32 doping levels. However, thermal decomposition of AgF$_2$ starts to occur at much lower temperature – ca. 960 K (690˚C) [25]. For this reason, such synthesis is unlikely to be successful. On the other hand, the enthalpies of formation of h-doped polymorphs are less unfavorable, and while they are expected to be more positive at higher temperatures, h-doped AgF$_2$ could in principle be obtained at higher F$_2$ pressures. Obviously, the opposite is expected to be true for e-doping.

The asymmetry between electron and hole doping in terms of energy cost is reflected in the body of knowledge about the Ag-F phase diagram, both experimental and computational. There are two known mixed-valent silver fluorides: Ag$_2$F$_5$ (or Ag$^{II}$F[Ag$^{III}$F$_4$]) [23] and Ag$_3$F$_8$ (or Ag$^{II}$[Ag$^{III}$F$_4$]$_2$) [26]. Both compounds contain Ag(II) and Ag(III) species and they exhibit mixed- and not intermediate-valence character, with different crystallographic sites for both types of silver cations. On the other hand, there are currently no known binary fluorides which would contain both Ag(I) and Ag(II) species, although at least one example of a ternary fluoride containing those species has been obtained. [27] Theoretical considerations regarding the possible stability of mixed-valent binary Ag(I)/Ag(II) fluorides have been previously published. [28] However, from the experience of our group and our collaborators, attempts at obtaining such compounds (which included controlled thermal decomposition of AgF$_2$) have in all but two instances led to separated phases of AgF and AgF$_2$. [29,30]

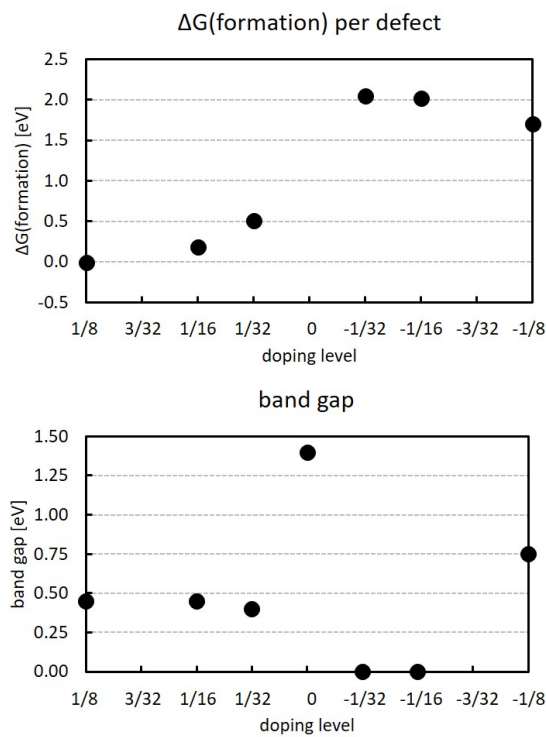

Fig. 5. Energy cost and band gap of doped $AgF_2$. Top panel: free energy of formation per defect. Bottom panel: band gap. On *x* axis, "+" represents hole doping and "-" – electron doping.

Given the insight from the structural data for undoped and doped polymorphs in this work, as well as the experimental data on silver fluorides [20,23,26,31], it can be noted that in terms of local coordination patterns, Ag(II) and Ag(III) species tend to be much more similar to each other than any of the two is to Ag(I). In other words, substitution of Ag(II) for Ag(III) exerts a much weaker strain on the crystal structure than its substitution for Ag(I) species. This may be the main reason for both (a) higher (more positive) energy of formation for e-doping in this work and (b) lack of Ag(I)/Ag(II) mixed valent binary fluorides in the experimental literature.

3. ELECTRONIC STRUCTURE

Fig. 6 shows the eDOS graphs of the obtained ground-state solutions for undoped and doped $AgF_2$. For the undoped, stoichiometric case, the band gap is predominantly between occupied ligand states of F and empty metal states of Ag, making this compound a classic charge-transfer insulator according to Zaanen-Sawatzky-Allen model [32], as has been shown in previous works. [1] There is also strong admixing between Ag and F states, which is attributed to strong covalence of Ag-F bonds in $AgF_2$, documented in previous X-ray photoelectron spectroscopy studies of silver fluorides. [33] Due to these characteristics, we can expect that additional occupied states in the e-doped solutions will have a majority Ag character, while holes will be largely of F character.

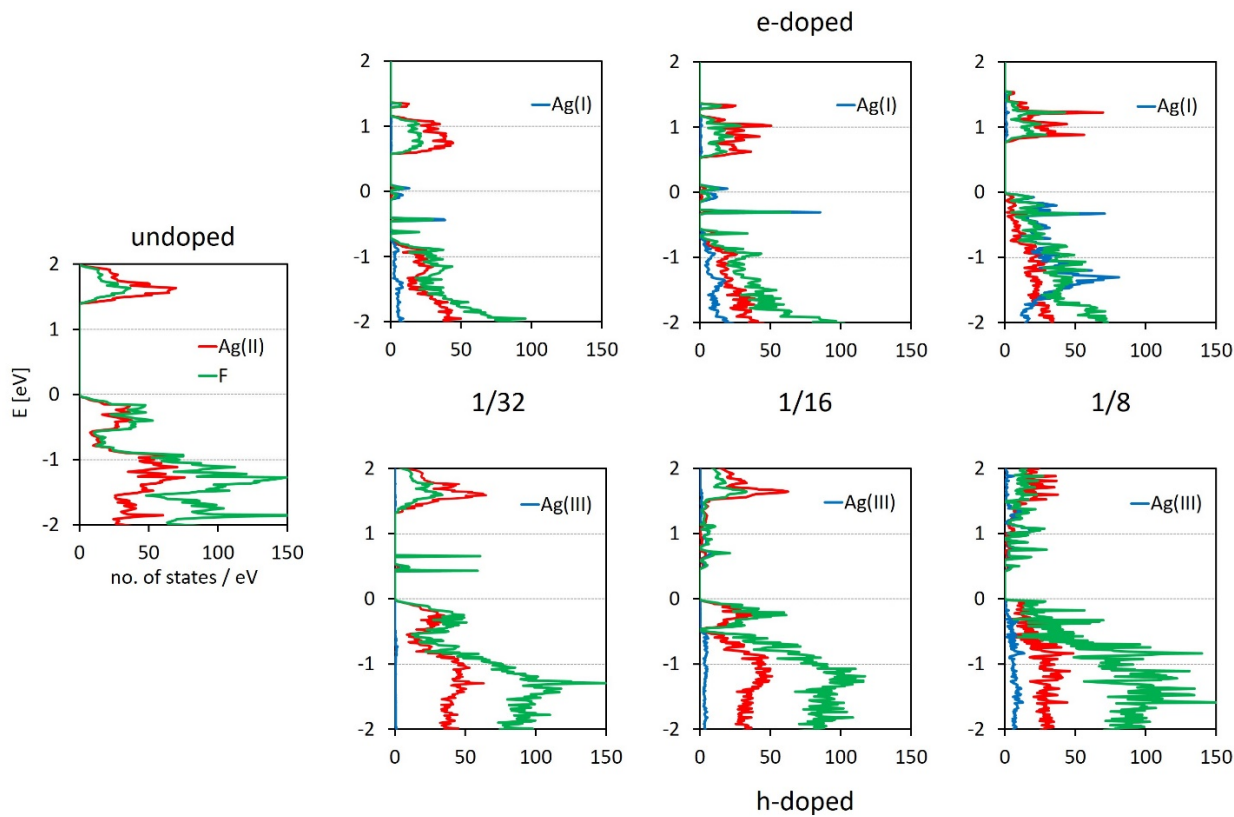

Fig. 6. eDOS plots for undoped AgF$_2$ and for doped AgF$_2$ polymorphs. Zero energy corresponds to E$_{Fermi}$.

Indeed, this is the case in the studied solutions. We used the previous insight from magnetic structure analysis to identify sites of Ag(II), Ag(I) and Ag(III) character in the electronic structure. The intermediate-spin Ag cations indicated in fig. 2 were designated as Ag(I) or Ag(III). As expected, the new states appear in the former band gap. For example, at 1/32 and 1/16 levels of e-doping, the new states are predominantly of the Ag(I) character and are less strongly bound, i.e. at higher energy in eDOS. These new bands are very narrow, in agreement with the presence of localized Ag(I) sites in the structure. Surprisingly, some of the new Ag(I) states appear just above the Fermi level and right next to Ag(I) states below it. These empty Ag(I) states are probably an artifact related to the previously discussed symmetrical distribution of additional spin as a superposition of two solutions, as shown in fig. 2. However, the band gap, while very small, retains a non-zero value – ca. 0.02 eV, which is smaller than the width of smearing used in eDOS calculations (0.15 eV). On the other hand, the new in-gap states in h-doped 1/32 and 1/16 models have a strong F contribution, which suggests that the holes are more localized on F than on Ag, and that the chemical species present in the structure is actually similar to F$^\bullet$ radical. At the same time, the Ag(III) states are much less pronounced. At 1/8 level, e-doping manifests as new Ag(I) states somewhat more dispersed just below the Fermi level, which also leads to a relative shift of empty Ag(II) bands and shrinking of the band gap to ca. 0.7 eV. In h-doped 1/8 solution, there are numerous, but narrow and localized states of both Ag(III) and F character (in roughly equal measure) within the former gap, effectively shrinking it to ca. 0.4 eV. Extended eDOS graphs can be found in ESI.

The picture emerging from the analysis of crystal, magnetic and electronic structures of doped AgF$_2$ polymorphs suggests a strong propensity of system towards self-trapping of defects introduced by modification of fluorine content. In particular, at the 1/8 doping level – which is close to the optimal level of doping in cuprate-based high-$T_c$ superconductors – localization of additional charge and reduction of band gap, compared to undoped AgF$_2$, are observed. This is similar, indeed, to so called 1/8-anomaly in cuprates. [34]

4. PRESSURE EFFECTS

As previously mentioned, pressure is also an interesting factor to study in the case of doped AgF$_2$. Increased pressure usually leads to broadening of electronic bands and eventually metallization, due to increased orbital overlap induced by shorter interatomic distances. [35] Pressure is also an important factor impacting the $T_c$ in oxygen-doped cuprates, which usually exhibit a maximum of $T_c$ in their phase diagram while at optimum external pressure. [36–39] Similar effects might be envisaged for the doped AgF$_2$. Since AgF$_2$ is known to undergo a drastic structural reorganization into a nanotubular polymorph at ca. 14 GPa [40], we decided to study only the range up to 10 GPa, as the second-order transition at ca. 7 GPa leads only to a slight symmetry lowering. [41] Selected data is presented for e-doping in fig. 7 and h-doping in fig. 8. In the case of e-doping, the increase of formation energy with pressure is apparent (Fig. 7, top), due to factors described above; namely, the equilibrium is shifted towards substrates, which consist only of AgF$_2$ and thus have lower volume (eq. 2). The band gap (Fig. 7, top) remains virtually unaffected for polymorphs at 1/8 and 1/32 levels of e-doping. However, the polymorph at 1/16 doping level undergoes an electronic and structural transition between 7.5 and 10.0 GPa into a polymorph similar to the one found at 1/8 doping level (Fig. 7, middle): the band gap becomes wide open again and equal to ca. 0.7 eV, and the Ag(I) cations are aligned in channels roughly along the *c* direction and with the distance to F atoms on average ca. 0.2 Å larger for the remaining Ag(II) cations. It appears, therefore, that doped AgF$_2$ resists metallization by rearranging the vacancies and opening the band gap, which is in line with generalized maximum hardness principle. [42] More detailed documentation of this process in e-doped AgF$_2$ can be found in ESI.

In the case of h-doped AgF$_2$, as predicted above, the free energy of formation mostly decreases with pressure (Fig. 8, top). The most noticeable process observed here is the transition of h-doped structure at 1/32 level between 7.5 and 10 GPa, which leads to redistribution of charge and spin: the single, intermediate-spin Ag site in the center of the unit cell is replaced by two high-spin Ag(III) cations – one in the center and another one on the *ab* wall of the unit cell (Fig.8, middle). As in the 1/8 h-doped solution, these high-spin Ag(III) species are coordinated by six F atoms at an average distance of 2.06 Å (compared to compressed octahedral pattern of intermediate-spin Ag center at lower pressure, see also table 1). This transition is most likely second-order, as there are no corresponding discontinuities in the pressure dependence of volume and lattice constants (cf. ESI). This process increases the band gap to ca. 0.25 eV, which was previously reduced in the 0-7.5 GPa range from 0.4 to 0.15 eV (Fig.8, bottom). A non-monotonic pressure dependence of band gap and free energy of formation is also notable for the 1/8 solution (Fig. 8, top). However, the dependence of cell volume and vectors is monotonic within the studied range and there are no discernible structural modifications. Most likely, this behavior stems from the presence of a variety of electronic states (high- and low-spin Ag(III), holes on F) in the first place, which respond to increasing pressure in different ways.

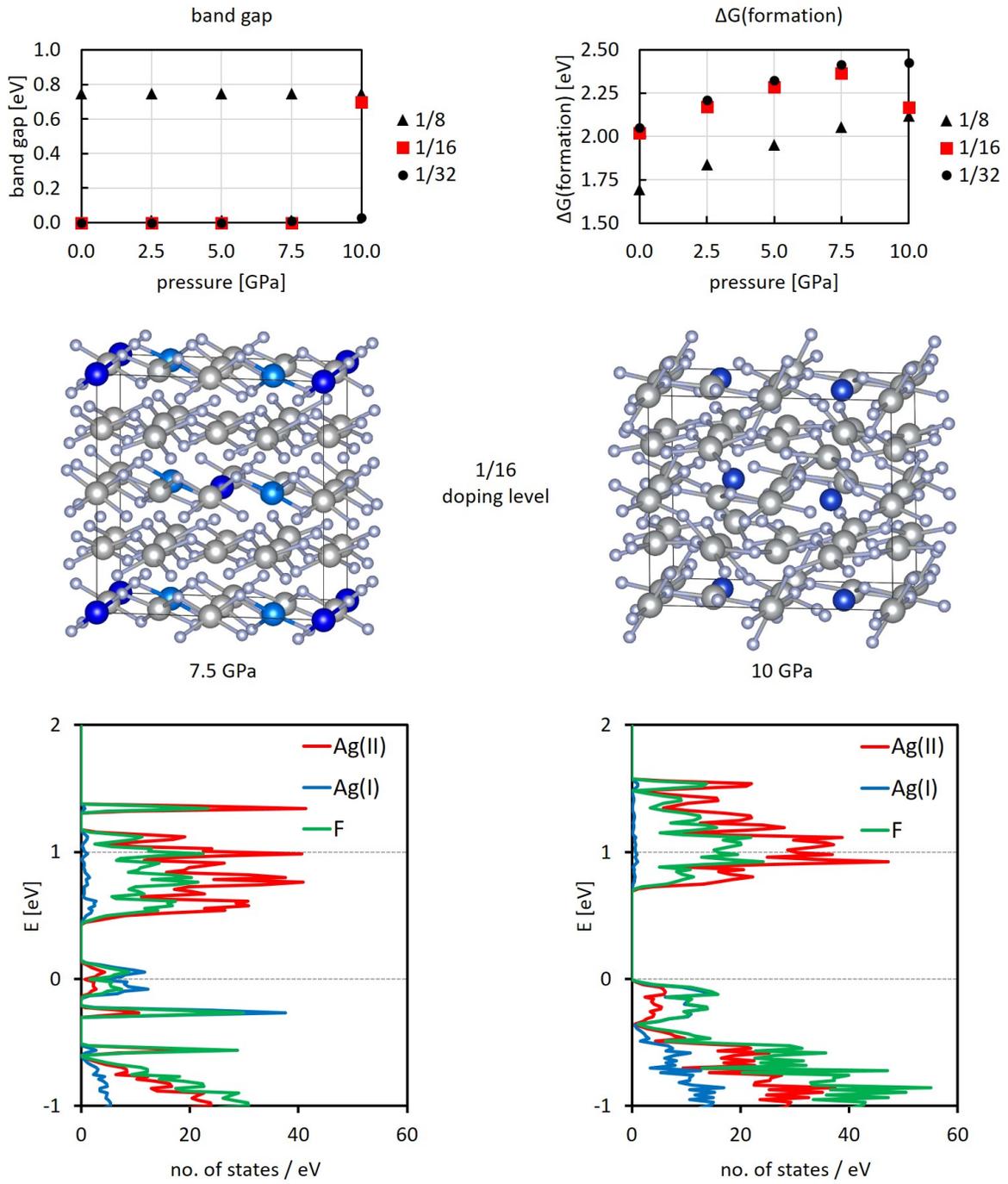

Fig.7. Pressure effects in e-doped AgF$_2$. Top panel: band gap and free energy of formation; middle panel: comparison of 1/16 e-doped structures before and after phase transition; bottom panel: eDOS graphs for 1/16 e-doped structures before and after phase transition.

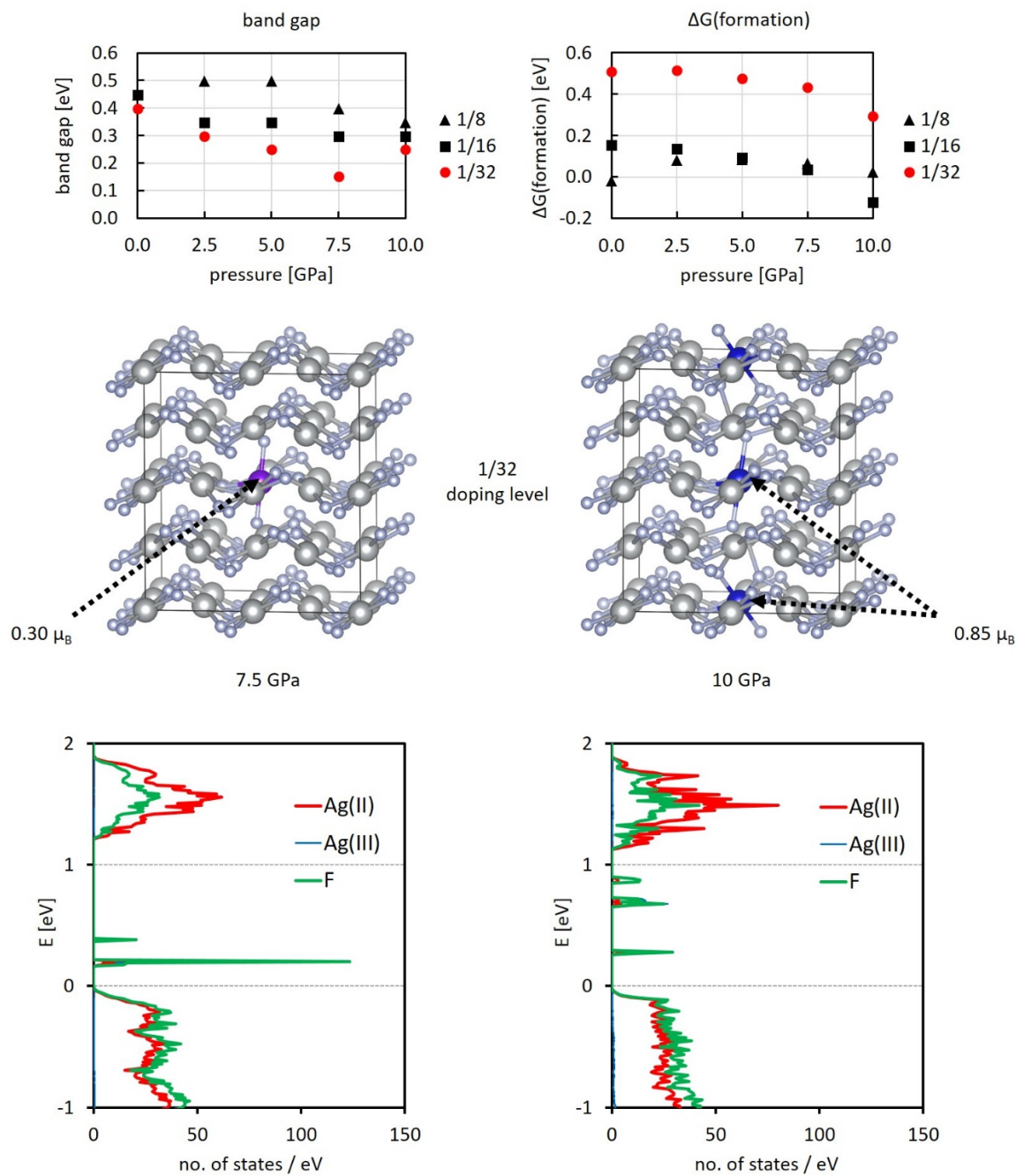

Fig. 8. Pressure effects in h-doped AgF$_2$. Top panel: band gap and free energy of formation; middle panel: comparison of 1/32 h-doped structures before and after phase transition; bottom panel: eDOS graphs for 1/32 h-doped structures before and after phase transition.

CONCLUSIONS

Our computational studies of chemical doping of AgF$_2$ indicate that: (a) electron- and hole-doped AgF$_2$ remains – for the most part – insulating, and resistant to metallization by pressure up to 10 GPa; (b) the additional electronic states arising upon doping are highly localized; (c) the spatial distribution of defects in the unit cell of lowest-energy solutions suggests propensity of the system towards phase separation.

Such results seem to indicate that chemical doping to $AgF_2$ is nearly impossible – as far as the conditions of thermodynamic equilibrium are assumed. In our recent study of polarons in $AgF_2$, lattice self-trapping of defects induced by modified electron count in the unit cell was also observed for bulk $AgF_2$ – but, importantly, to a lesser extent in hypothetical polymorphs with flat monolayers. [6] This is also in line with strong covalence of Ag-F bonds, well-documented in previous studies, [33] which leads to strong vibronic coupling and in turn increases the tendency for trapping of defects. Another thing to consider is that the type of doping studied here, while in principle similar to chemical modifications of cuprates, is by definition more local: the fluorine defects/interstitials ultimately have to occupy a particular position in the unit cell, different than any of the original F atoms in the stoichiometric $AgF_2$ because of symmetry constraints (or rather, the appearance of said modifications breaks some of those constraints). In cuprates, doping is usually achieved by substitution of a fraction of closed-shell cations for other with different oxidation state, and while this is often associated with oxygen deficiency, these modifications ultimately exert a much weaker structural influence on the resulting compound. $AgF_2$, while a direct analog to $La_2CuO_4$ in many ways, is a binary rather than ternary compound, which makes such substitutions at the La site impossible. Given all of the above, chemical doping of $AgF_2$, even combined with application of external pressure, is unlikely to produce materials with properties that could be promising from the point of view of high-temperature superconductivity. There are, however, different pathways towards cuprate-like properties in $AgF_2$ that are currently being explored – in particular, a flat $AgF_2$ monolayer stabilized and doped by using an appropriate substrate for epitaxy. [7] Moreover, one may search for doped $AgF_2$ in the conditions far from the thermodynamic equilibrium; indeed, two of the higher-energy structures found in this work are predicted to exhibit metallic DOS (section S4 in ESI).

ACKNOWLEDGEMENTS

WG thanks the Polish National Science Center (NCN) for the Maestro project (2017/26/A/ST5/00570). This research was carried out with the support of the Interdisciplinary Centre for Mathematical and Computational Modelling (ICM), University of Warsaw under grant ADVANCE++ (no. GA76-19). MD acknowledges the ERDF fund, Research and Innovation Operational Programme (ITMS2014+: 313011W085), Scientific Grant Agency of the Slovak Republic (VG 1/0223/19) and the Slovak Research and Development Agency (APVV-18-0168).

# Defect trapping and phase separation in chemically doped bulk AgF$_2$

Adam Grzelak[1]*, Mariana Derzsi[1,2], Wojciech Grochala[1]*

[1]Center of New Technologies, University of Warsaw, Banacha 2C, 02-097 Warsaw, Poland
[2]Advanced Technologies Research Institute, Faculty of Materials Science and Technology in Trnava, Slovak University of Technology in Bratislava, Jána Bottu 8857/25, 917 24 Trnava, Slovakia

SUPPLEMENTARY INFORMATION

S1. Expanded eDOS graphs

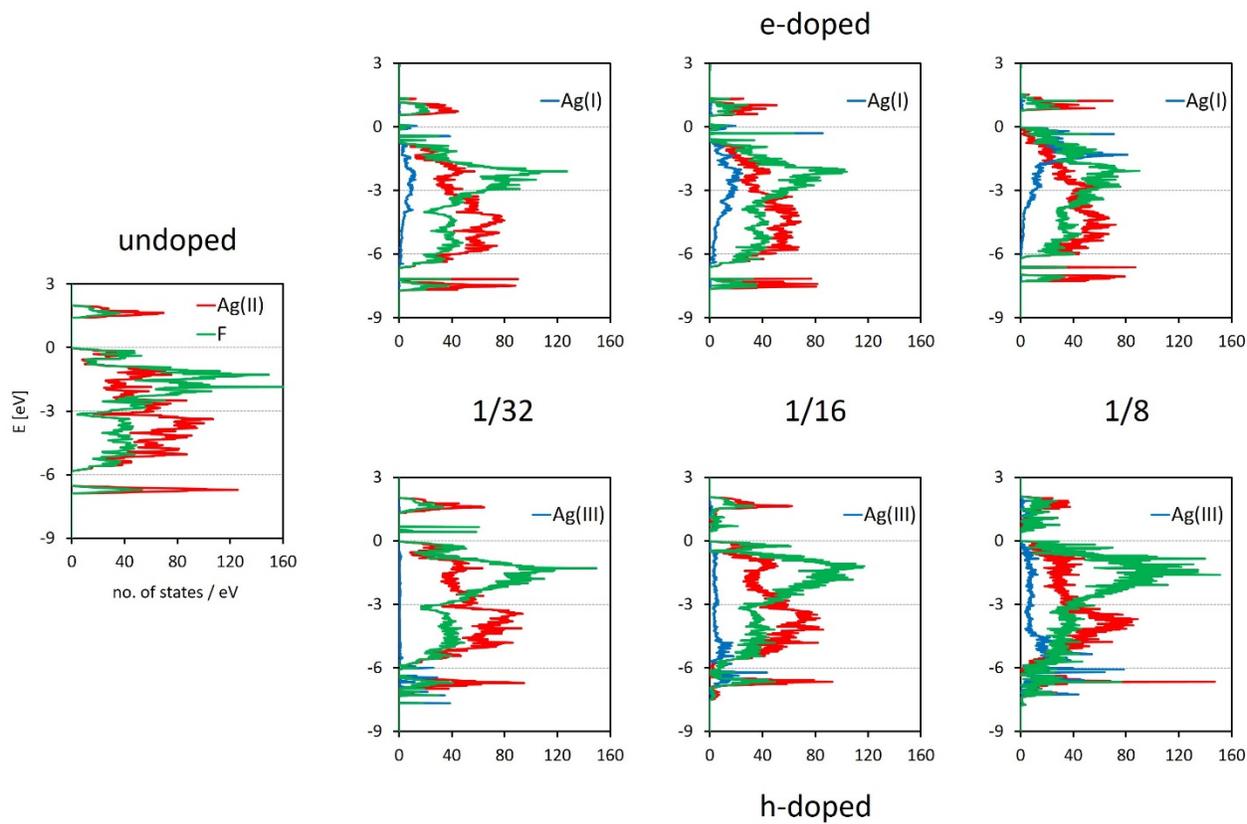

Fig. S1.1. eDOS plots for undoped AgF$_2$ and for doped AgF$_2$ polymorphs. Zero energy corresponds to E$_{Fermi}$.

## S2. Pressure dependence of lattice constants for doped AgF$_2$ polymorphs

**undoped AgF$_2$ (disregarding known phase transitions)**

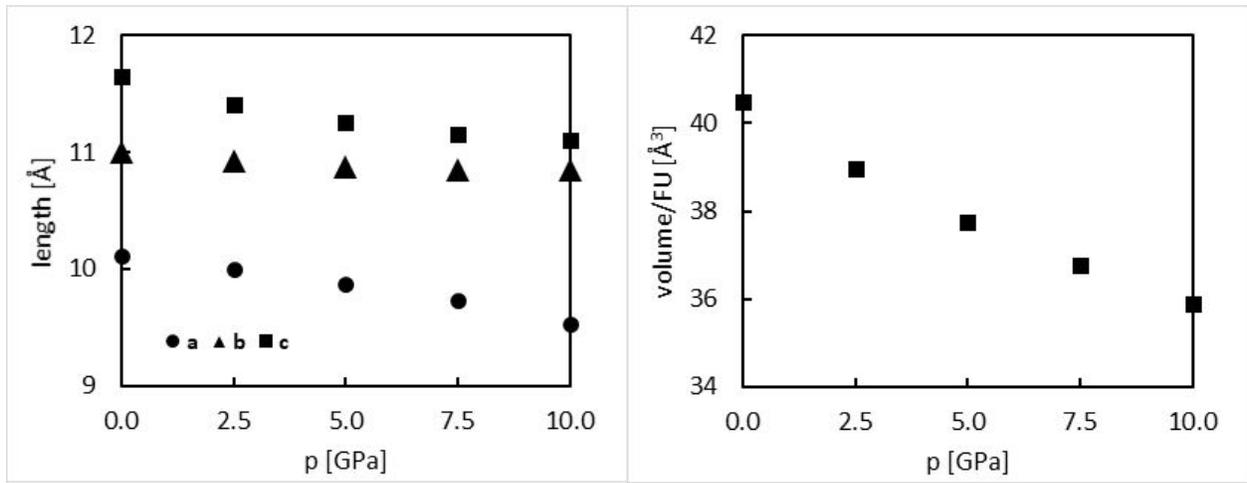

**1/32 e-doping**

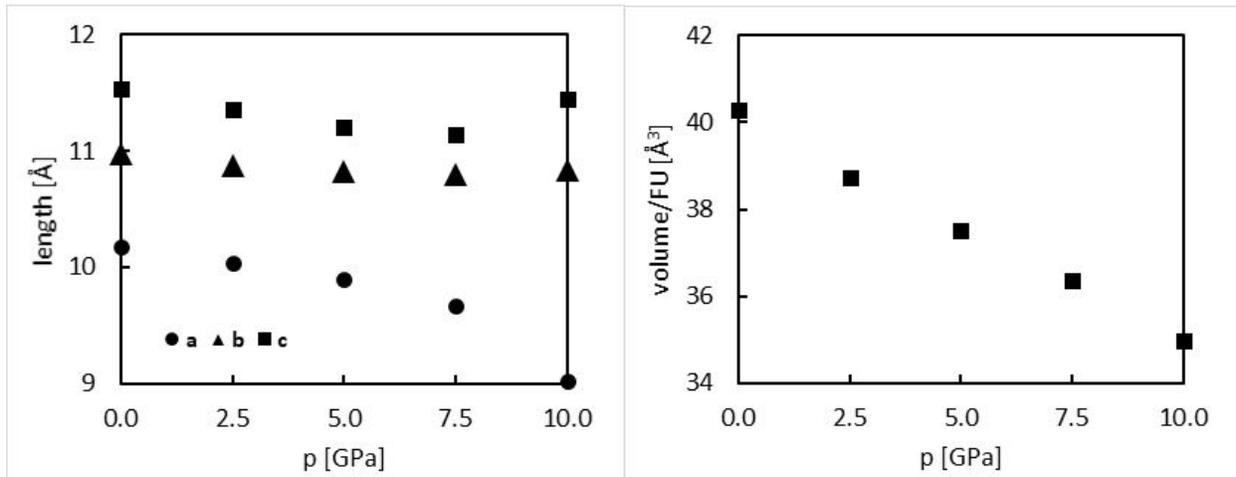

**1/16 e-doping**

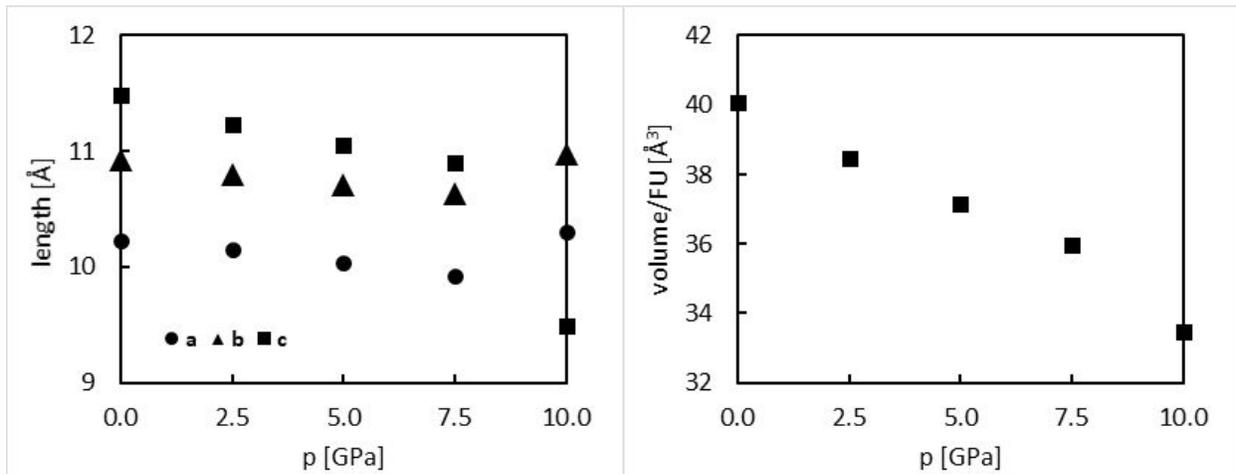

**1/8 e-doping**

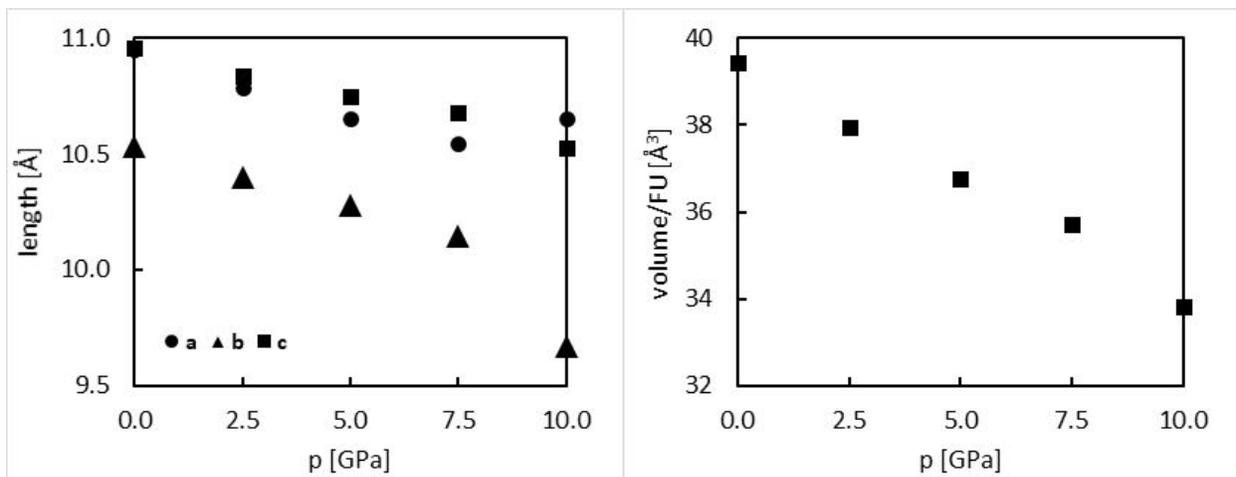

**1/32 h-doping**

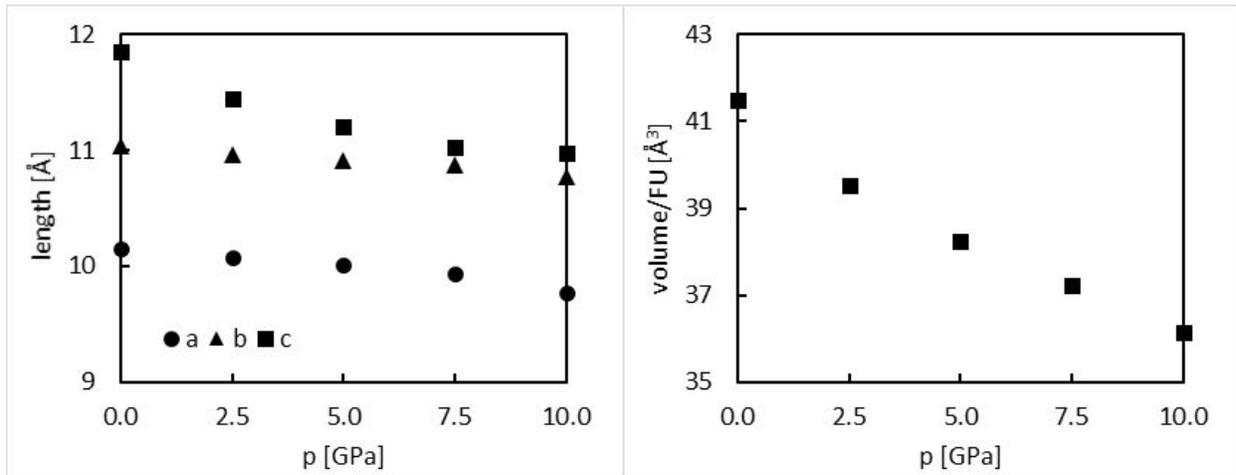

**1/16 h-doping**

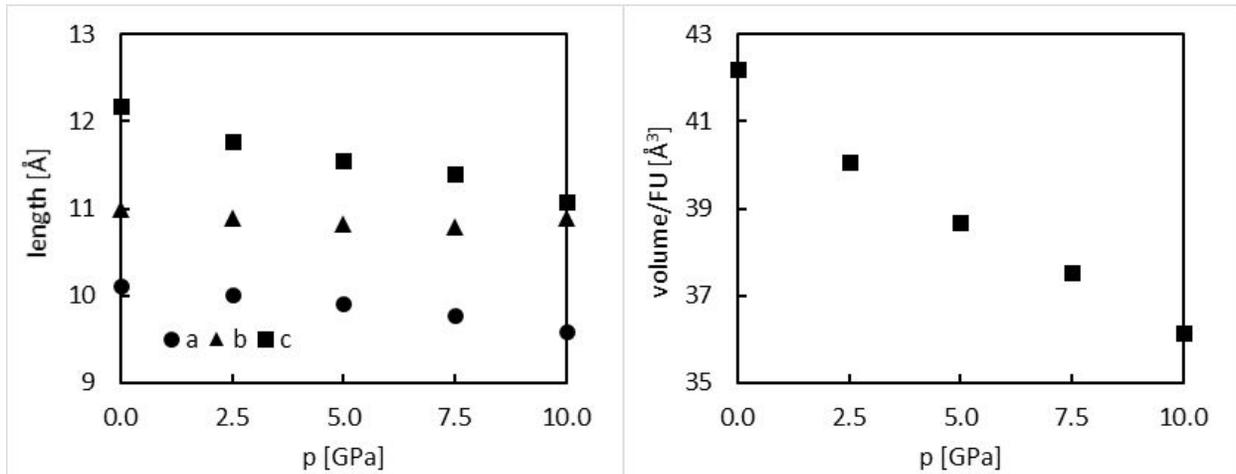

**1/8 h-doping**

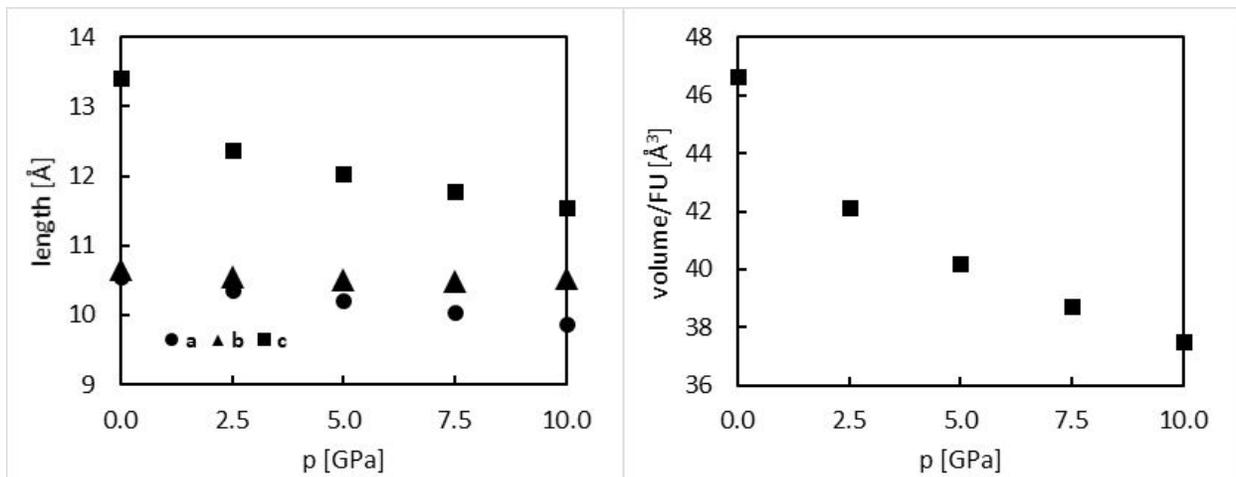

## S3. Resistance of e-doped AgF$_2$ to metallization

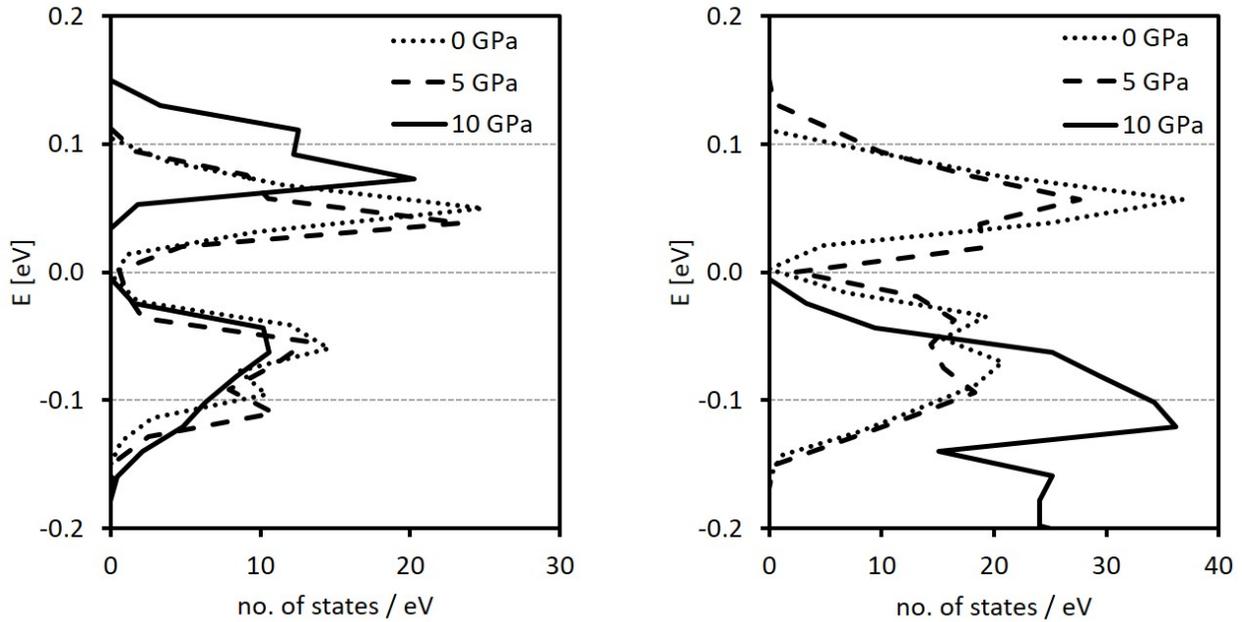

Fig. S3.1. Pressure dependence of unit cell vectors of e-doped polymorph of AgF$_2$.

In e-doped AgF$_2$ at 1/32 and 1/16 level, there exist additional states within the former band gap and right next to the new Fermi level. While at 0 GPa in both polymorph the gap is very small but not null, compression leads to greater orbital overlap and subsequent metallization. In 1/32 polymorph, the system counteracts this tendency by slight elongation of *b* and *c* vector and stronger compression in *a* direction, which is enough to open the gap again at 10 GPa. In the 1/16 case, however, there are much more of these new states and the system is more prone to metallize, therefore a phase transition (shown in main text) is necessary to counteract this process.

## S4. Selected higher-energy solutions

We present here two alternative models found in our study, which exhibit non-zero density of states at the Fermi level and are therefore weakly metallic. Both of these solutions are ca. 0.05 eV (per AgF$_2$ formula unit) higher in energy than the ground-state solutions presented in the main paper. Magnetic structure and electronic density of states are shown for the two cases. In both cases there is additional spin localized on some of the F atoms, but this is not indicated for clarity. For the eDOS calculations, both null-spin blue Ag(I) centers as well as intermediate-spin (IS) Ag centers are considered as contributing to Ag(I) states.

If those or similar metastable structures could be obtained, they might exhibit metallic character and possibly also superconductivity.

**1/16 e-doping – alternative solution**

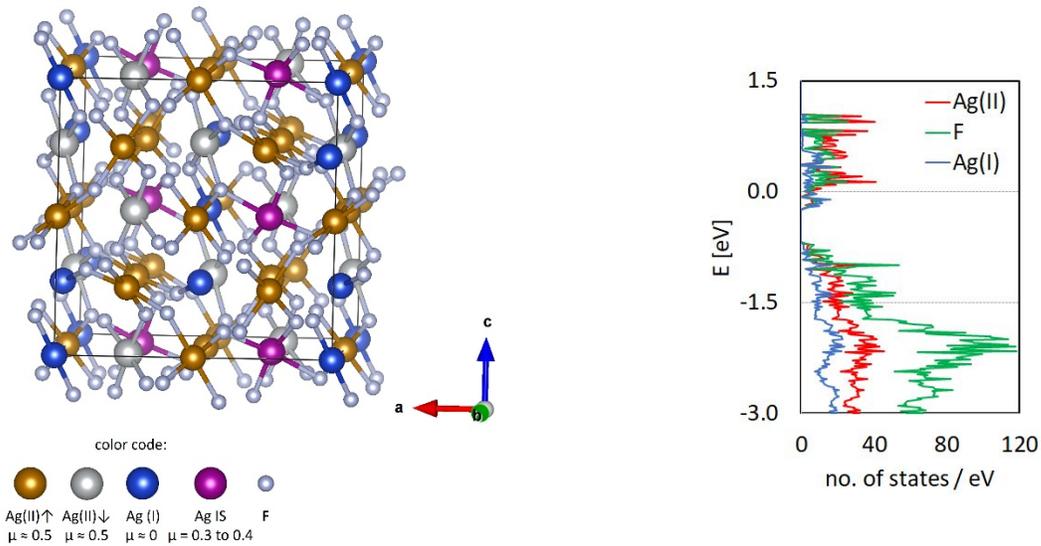

color code:
Ag(II)↑  Ag(II)↓  Ag(I)  Ag IS  F
μ ≈ 0.5  μ ≈ 0.5  μ ≈ 0  μ = 0.3 to 0.4

**1/8 e-doping – alternative solution**

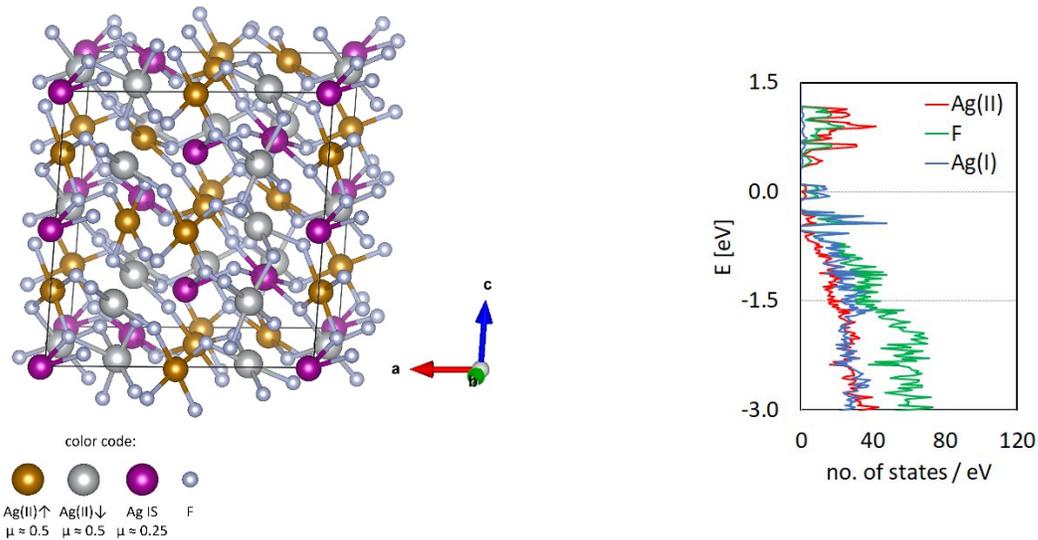

color code:
Ag(II)↑  Ag(II)↓  Ag IS  F
μ ≈ 0.5  μ ≈ 0.5  μ ≈ 0.25